\documentclass[10pt]{amsart}
\usepackage[T1]{fontenc}
\usepackage[centertags]{amsmath}
\usepackage{amsfonts}
\usepackage{amssymb}
\usepackage{amsthm}
\usepackage{newlfont}
\usepackage{epsfig}
\usepackage{amscd}

\newcommand{\CC}{{\mathbb C}}

\newcommand{\ZZ}{{\mathbb Z}}
\newcommand{\KK}{{\mathbb K}}
\newcommand{\FF}{{\mathbb F}}

\newcommand{\calC}{{\mathcal C}}

\newtheorem{Th}{Theorem}

\newtheorem{Lem}[Th]{Lemma}
\newtheorem{Prop}[Th]{Proposition}
\theoremstyle{definition}
\newtheorem{Def}{Definition}
\theoremstyle{remark}
\newtheorem*{Rem}{Remark}
\begin{document}

\title[Integrable KP and KdV cellular automata]{Integrable KP and K{d}V cellular automata \\
 out of a hyperelliptic curve}
\author[M. Bia{\l}ecki]{Mariusz Bia{\l}ecki}
\address{M. Bia{\l}ecki \newline 
Institute of Geophysics, Polish Academy of Sciences\\ ul. Ksi\c{e}cia Janusza 64\\ 01-452 Warszawa\\ 
Poland\\ \newline
Institute of Theoretical Physics, University of Bialystok\\ ul. Lipowa 41\\ 15-424 Bia{\l}ystok\\ 
Poland}
\email{bialecki@igf.edu.pl} 

\begin{abstract}

\noindent The goal of this paper is 
to present a solution of the cellular automaton associated 
with the discrete KdV equation using
algebro-geometric solution
of the discrete KP equation over a finite field  
out of a hyperelliptic curve.

\bigskip

\end{abstract}

\maketitle

\section{Introduction}

Cellular automata (CA) are (solutions of) completely discrete  
dynamical systems, for which values of both independent and dependent variables are discrete. 
The concept of CA emerged in interaction with computational machines and hence 
CA are convenient tools for computer simulations of various phenomena.
There is also an approach to CA, in which the main 
objective is to obtain "analitic" solutions (or full solutions or in a form  
good for investigation of their global properties) without need of performing step by step
calculations. This leads naturally to notion of integrability and 
integrable cellular automata (ICA). (For references to ICA see \cite{BD-KP}.)

Recently a new method of construction of ICA was proposed in \cite{DBK}.
Its main idea is to keep the form of a given
integrable discrete system and to transfer the algebro-geometric method of 
construction of its solutions~\cite{Krich-discr,BBEIM}
from the complex field $\CC$ to a finite field case. 
In this framework there were constructed finite field versions of 
multisoliton solutions for the fully discrete 2D Toda system (the Hirota equation) in \cite{DBK}
and for discrete KP and KdV equations (in Hirota form) in \cite{BD-KP} and also 
algebro-geometric solutions of the discrete KP equation out of a hyperelliptic 
curve in \cite{BD-hyp}.

In this paper we extend previous works to construct algebro-geometric solution
of the discrete KdV equation out of a hyperelliptic curve. To reach this aim 
we obtain a solution of dKP equation in a form 
compatibile with the reduction from the dKP equation to the dKdV equation.
  
The paper is constructed as follows. In Section \ref{sec:ffKP-alg-geom} we 
first summarize the finite field version of the algebro-geometric construction of
solutions of the discrete KP and KdV equations. In Section \ref{sec:hyp}
we apply the method to construct solution of the discrete KP and KdV equations
starting from an algebraic curve of genus two.

\section{The finite field solution of the discrete KP equation out of nonsingular
algebraic curves} \label{sec:ffKP-alg-geom}
We first shortly recall algebro-geometric construction of solutions
of the discrete KP equation over finite fields and its reduction 
to the discrete KdV equation~\cite{DBK,BD-KP,BD-hyp}. 
An algebro-geometric approach in case of complex field $\CC$ is described in \cite{BBEIM}. 
An algebraic curves over finite fields are exposed for example in \cite{Moreno,Sticht}.

\subsection{General construction for the dKP equation} \label{sec:gen-constr}
For the general construction we need an algebraic projective curve $\calC/\KK$ (or simply $\calC$),  
absolutely irreducible, nonsingular, of genus $g$, defined over the finite  
field $\KK=\FF_q$ with $q$ elements.
By $\calC(\KK)$ we denote 
the set of $\KK$-rational points of the curve. 
By $\overline{\KK}$ denote the algebraic closure of  
$\KK$, i.e., $\overline{\KK} = \bigcup_{\ell=1}^\infty \FF_{q^\ell}$, and by 
$\calC(\overline{\KK})$ denote the corresponding infinite set of 
$\overline{\KK}$-rational points of the curve. 
Denote by $\mathrm{Div}(\calC)$ the abelian group of the divisors on the curve $\calC$.
The action of the Galois group $G(\overline{\KK}/\KK)$ (of automorphisms of  
$\overline{\KK}$ which are identity on $\KK$) extends 
naturally to action on $\calC(\overline{\KK})$ and $\mathrm{Div}(\calC)$. 
A field of $\KK$-rational functions on the curve $\calC$ we denote by $\KK(\calC)$ 
and the vector space $L(D)$ is defined as $\{ f \in \overline\KK(\calC) \ | \ (f) > -D \}$,
where $D \in \mathrm{Div}(\calC)$ and $(f)=\sum_{P\in \calC} ord_P (f)\cdot P$ \ 
is the divisor of the function $f\in\KK(\calC)$.

On the curve $\calC$ we choose:
\begin{enumerate} 
\item four points $A_i\in\calC(\KK)$, \  $i=0,1,2,3$,  
\item effective $\KK$-rational  divisor of order $g$,
i.e., $g$ points $B_\gamma\in\calC(\overline{\KK})$, $\gamma=1,\dots,g$, 
which satisfy the following $\KK$-rationality condition 
\[ 
\forall \sigma\in  G(\overline{\KK}/\KK), \quad  
\sigma(B_\gamma) = B_{\gamma^\prime}.
\] 
\end{enumerate}
We assume that all the points used 
are distinct and in general position. In particular, 
the divisor $\sum_{\gamma=1}^g B_\gamma$ (and a divisor $D(n_1,n_2,n_3)$ defined below) is non-special. 
\begin{Def}  \label{def:psi}
Fix $\KK$-rational local parameter $t_0$ at $A_0$. 
For any integers $n_1,n_2,n_3\in \ZZ$ let divisor $D(n_1,n_2,n_3)$ be of the form
$$D(n_1,n_2,n_3)=n_1(A_0-A_1)+n_2(A_0-A_2)+n_3(A_0-A_3) + \sum_{\gamma=1}^g B_{\gamma}.$$
The function  
$\psi(n_1,n_2,n_3)$ (called a wave function) is a 
rational function on the curve $\calC$ with the following properties
\begin{enumerate}
\item the divisor of the function satisfy $(\psi)> -D$, i.e. $\psi \in L(D)$,  
\item the first nontrivial coefficient of its expansion in $t_0$ at $A_0$ is 
normalized to one. \label{point-norm}  
\end{enumerate}
\end{Def} 
Existence and uniqueness of the function $\psi(n_1,n_2,n_3)$ is due 
to application of the Riemann--Roch theorem with general position
assumption and due to normalization. 
Moreover, the function $\psi(n_1,n_2,n_3)$ is $\KK$-rational, which follows  
from $\KK$-rationality conditions for sets of points in their definition. 

\begin{Rem}
Notice that the function $\psi(n_1,n_2,n_3)$ has $g$ zeros not {\bf explicitely} specified
in the Definition~\ref{def:psi}. 
\end{Rem}

The next step of the construction is to obtain linear equations
for wave functions. The full form of such equation is in case when 
the pole of $\psi(n_1,n_2,n_3)$ at $A_0$ is of the order exactly $(n_1+n_2+n_3)$ 
and respective zeros at $A_i$ are of the order $n_i$, for $i=1,2,3$. 
We will call this case generic.
Having fixed $\KK$-rational local parameters $t_i$ at $A_i$, $i=1,2,3$, 
denote by $\zeta^{(i)}_k(n_1,n_2,n_3)$, $i=1,2,3$, 
the $\KK$-rational
coefficients of expansion of $\psi(n_1,n_2,n_3)$ at $A_i$, respectively, 
i.e.,
\begin{align*} 
\psi (n_1,n_2,n_3) &= t_i^{n_i} \sum_{k=0}^{\infty} 
\zeta^{(i)}_k(n_1,n_2,n_3)  t_i^k , \quad i=1,2,3.  
\end{align*} 
Denote by $T_i$ the operator of translation 
in the variable $n_i$, $i=1,2,3$, 
for example $T_2 \psi(n_1,n_2,n_3) = \psi(n_1,n_2+1,n_3) $. 
The full linear equation is of the form
 \begin{equation}  
T_i \psi - T_j \psi +  
\frac{T_j\zeta^{(i)}_ 0}{\zeta^{(i)}_0} \psi = 0 , \quad i\ne j , 
\quad i,j=1,2,3. 
\label{eq:psi}  
\end{equation} 
It follows from observation, that $T_i \psi - T_j \psi \in L(D)$, hence must be
proportional to wave function $\psi$. Coefficients of proportionality can be 
obtained from comparison (the lowest degree terms) of expansions of left and 
right sides of \eqref{eq:psi} at the point $A_i$.

\begin{Rem}
When the genericity assumption fails then the linear problem~\eqref{eq:psi} 
degenerates to the form $T_i\psi=\psi$ or even to $0=0$. 
\end{Rem}

Notice that equation \eqref{eq:psi} gives
\begin{equation} \label{eq:zeta-zeta}
\frac{T_j\zeta^{(i)}_ 0}{\zeta^{(i)}_0} = -  
\frac{T_i\zeta^{(j)}_ 0}{\zeta^{(j)}_0} , \quad i\ne j,\quad i,j=1,2,3.  
\end{equation}
Define 
\begin{equation} \label{eq:rho-def}
\rho_i= (-1)^{\sum_{j<i} n_j} \zeta^{(i)}_0, \quad i=1,2,3,
\end{equation} 
then equation
\eqref{eq:zeta-zeta} implies existence of a $\KK$-valued potential 
(the $\tau$-function) defined (up to a multiplicative
constant) by formulas 
\begin{equation} \label{eq:tau-def} 
 \frac{T_i\tau}{\tau} = \rho_i,\quad  i=1,2,3.  
\end{equation} 
Finally, equations \eqref{eq:psi} give rise to condition 
\begin{equation}  \label{eq:KP-rho}
\frac{T_2\rho_1}{\rho_1} - \frac{T_3\rho_1}{\rho_1} + 
\frac{T_3\rho_2}{\rho_2} = 0, 
\end{equation}  
which written in terms of the $\tau$-function gives the discrete KP
equation~\cite{Hirota} called also the Hirota equation
\begin{equation}  \label{eq:tauKP} 
(T_1\tau) \;(T_2T_3\tau) - (T_2\tau) \;(T_3T_1 \tau) + 
(T_3\tau) \;(T_1T_2\tau) = 0. 
\end{equation} 
\begin{Rem} 
Equation \eqref{eq:KP-rho} can be obtained also from expansion of equation 
\eqref{eq:psi} at $A_k$, where $k=1,2,3$, $k\ne i,j$.
\end{Rem}

Absence of a term in the linear problem \eqref{eq:psi} reflects, due to
Remark above, in absence of
the corresponding term in equation \eqref{eq:tauKP}. This implies that in the
non-generic case, when we have not defined the $\tau$-function yet, we are 
forced to put it to zero. 

\subsection{Reduction to the dKdV equation}
\label{sec:alg-dKdV}
The discrete KdV equation \cite{HirotaKdV,KWZ}
\begin{equation}  \label{eq:tauKdV} 
(T_1\tau) \;\tau - (T_3^{-1}\tau) \;(T_3T_1 \tau) + 
(T_3\tau) \;(T_1T_3^{-1}\tau) = 0, 
\end{equation} 
is obtained from the discrete KP equation by imposing
constraint
\begin{equation} \label{eq:KPtoKdVconstr}
T_2T_3 \tau = \gamma \tau,
\end{equation}
where $\gamma$ is a non-zero constant.
Algebro-geometric solutions of the discrete KdV equation can be constructed 
using following facts (see \cite{BD-KP}).
\begin{Lem} \label{lem:KP-to-KdV}
Assume that on the algebraic curve $\calC$ there exists a rational
function $h$ with the following properties
\begin{enumerate}
\item the divisor of the function is $(h)=A_2+A_3-2A_0$,
\item the first nontrivial coefficient of 
its expansion in the parameter $t_0$ at $A_0$ is normalized to one.
\end{enumerate}
Then the wave function $\psi$ satisfies the following condition
\begin{equation} \label{eq:KPtoKdV-constr-psi}
T_2T_3\psi = h \psi.
\end{equation}
\end{Lem}
\begin{Rem}
Existence of such a function $h$ implies that the algebraic curve $\calC$ is
hyperelliptic.
\end{Rem}
\begin{Prop} \label{prop:KP-to-KdV}
Let $h$ be the function as in Lemma~\ref{lem:KP-to-KdV}. Assume
additionally that
\begin{equation} \label{eq:assumpt-ha1}
h(A_1)=1.
\end{equation} 
Denote by $\delta_2$ and $\delta_3$ the respective first 
coefficients of local
expansion of $h$ in parameters $t_2$ and $t_3$ at $A_2$ and $A_3$, 
i.e.
$h=t_2(\delta_2 + \dots), \quad h=t_3(\delta_3 + \dots)$.
Then the function
\begin{equation} \label{eq:tau-tau-tylda}
\tilde\tau= \tau \, \delta_2^{-n_2(n_2-1)/2}(-\delta_3)^{-n_3(n_3-1)/2}
\end{equation}
satisfies the discrete KdV equation \eqref{eq:tauKdV}.
\end{Prop}

\section{A "hyperelliptic" solution of the discrete KP and KdV equation}
\label{sec:hyp}

Our goal here is to construct a solution of the dKP equation to which 
we can apply the reduction scheme described above. 
We are forced to perform steps of the construction (see also \cite{Bial-phd} for details)
starting from a hyperelliptic curve. In detail we deal with a curve of genus $g=2$
but the technical tools used here can be applied directly to 
hyperelliptic curves of arbitrary genus.

\subsection{Hyperelliptic curves and Jacobian picture of the construction}
In the following we use an affine picture of a hyperelliptic curve.
It is motivated by the fact that general hyperelliptic curve can be tranformed to 
the form with only one point at infinity (see~\cite{Shafarevich}). 

\begin{Def}
A hyperelliptic curve $\calC$ of the genus $g$ over a field $\KK$ 
is given by 
\begin{equation} \label{eqn:hypdef}
 \calC: \quad v^2+ h(u)v-f(u)=0,
\end{equation}
where $h(u)\in\KK[u]$ is a polynomialof degree at most $g$, 
$f(u)\in\KK[u]$ is a monic polynomial of degree $2g+1$,
if there is no points $(u=x,v=y)\in \overline \KK \times \overline \KK$ which
satisfy equation~\eqref{eqn:hypdef} and equations $2y+h(x)=0$ and $h'(x)y-f'(x)=0$.
\end{Def}

By $\tilde P$ we denote the point opposite to $P$, i.e. 
conjugate with respect to hyperelliptic isomorphism.
Denote by $\mathrm{Div}^0(\calC;\KK)$ the abelian group of the
$\KK$-rational divisors on the curve $\calC$
and by 
$J(\calC;\KK)$ the group of eqivalence classes of $\KK$-rational 
degree zero divisors $\mathrm{Div}^0(\calC;\KK)$   modulo the $\KK$-rational 
principal divisors, i.e. divisors of a functions $\KK(\calC)$.
(In terms of algebraic geometry $J(\calC;\KK)$ is identified with the group of $\KK$-rational
points of the Jacobian of the curve $\calC$ ~\cite{Lang,Milne}.)

Two divisors $A,B \in \mathrm{Div}^0(\calC;\KK)$ are equivalent (we write $A \sim B$) 
if $B=A+(f)$ for some function $f\in\KK(\calC)$. The class of a divisor $A$ in the divisor class
group $J(C;\KK)$ is denoted by $[A]$.  
For a hyperelliptic curve $\calC$ (of genus $g$) each class of equivalence $[A]\in J(C;\KK)$ 
has a unique representant of the form of reduced divisor~\cite{hiper-dodatekAng}. 
\begin{Def}
A divisor $D\in \mathrm{Div}^0(\calC;\KK)$ of the form 
\begin{equation*}
D=\sum_{\gamma=1}^k  X_\gamma - k \cdot A_0, 
\end{equation*}
where $X_\gamma \in \calC \setminus \{A_0\}$ is called reduced if 
\begin{enumerate}
\item $k \leq g$
\item $\tilde X_{\gamma} \neq X_{\gamma'}$ \quad for all $\gamma \neq \gamma'$
\end{enumerate}
\end{Def}

Let us present in this picture the description of the wave function $\psi$ and
of the $\tau$-function. Consider the following divisor 
$D(n_1,n_2,n_3)\in\mathrm{Div}^0(\calC;\KK)$ of degree zero
\[ D(n_1,n_2,n_3)=n_1(A_0-A_1)+ n_2(A_0-A_2)+ n_3(A_0-A_3) +
\sum_{\gamma=1}^g B_\gamma - g\cdot A_0,
\]
with linear dependence on $n_1$, $n_2$ and $n_3$.
Its equivalence class in $\mathrm{J}(\calC;\KK)$ has 
the unique $\KK$-rational representant
of the form of reduced divisor
\[ X(n_1,n_2,n_3) = \sum_{\gamma=1}^k X_\gamma(n_1,n_2,n_3) - k\cdot A_0 .
\] 
This equivalence is given by a function 
whose divisor  is
\[
n_1(A_1-A_0)+ n_2(A_2-A_0)+ n_3(A_3-A_0) +
\sum_{\gamma=1}^k X_\gamma (n_1,n_2,n_3) -
\sum_{\gamma=1}^g B_\gamma + (g-k)A_0 .
\]
If we normalize such a function at $A_0$ acording to
Definition~\ref{def:psi} it becomes the wave function $\psi$. 
Notice that if $k \neq g$, it means that
the pole of the wave function at $A_0$ is of the order less then $(n_1+n_2+n_3)$, 
and it is a non-generic case, thus 
$\tau(n_1,n_2,n_3)=0$.

\begin{Rem}
Points $X_\gamma$, indicate 
zeros of
the wave function which are not explicitely specified in the previous construction. 
\end{Rem}

\subsection{A curve and its Jacobian}
Consider a hyperelliptic curve $\calC$ of genus $g=2$,  
defined over the field $ \FF_7 $ and given by the equation
\begin{equation}\label{eq:curve} 
\calC: \quad v^2 + u v = u^5 + 5 u^4 + 6 u^2 + u + 3.
\end{equation}
The $(u,v)$ coordinates of its  $\FF_7$-rational points are presented in 
Table~\ref{tab:PunktyF7}. 

The only two special points of the curve are
$(6,4)$ and the infinity point $\infty$. 

\begin{table}
\begin{center}
\begin{tabular}{|c||c|c|}
\hline \hline
$i$ &  $ P_i$  & $\tilde P_i$  \\
\hline \hline
0 &   $\infty$   & $P_0 $\\
1 &   $(1,1)$    & $(1,5)$    \\
2 &   $(2,2)$    & $(2,3)$    \\
3 &   $(5,3)$    & $(5,6)$    \\
4 &   $(6,4)$    & $P_4$ \\
\hline \hline
\end{tabular}
\bigskip
\caption{$\FF_7$-rational points of the curve $\calC$.
The  point opposite to $P$ (conjugate with respect to the hyperelliptic automorphism)
 is denoted by $\tilde P$.}
\label{tab:PunktyF7}
\end{center}
\end{table}
We identify the field $\FF_{49}$ with the extension of $\FF_7$ by the polynomial
$x^2+2$, i.e., $\FF_{49}={\FF_7[x]}/(x^2+2)$. 
It is convenient to introduce the following notation: the element $k \in \FF_{49}$ 
represented by the polynomial $ \beta x + \alpha $ is
denoted by the {\em natural} number $7 \beta + \alpha $. 
The Galois group $G(\FF_{49}/\FF_7)=\{ id, \sigma  \}$, where 
$\sigma$ is the Frobenius automorphism, acts  
on elements of $\FF_{49}\setminus\FF_7$ in the following way:
$ k = 7 \beta +\alpha \mapsto \sigma(k) = 7(7-\beta )+\alpha $.
The coordinates of
$\FF_{49}$-rational points of the curve (which are not $\FF_7$-rational) 
are presented in Table~\ref{tab:PunktyF49}. 

\begin{table}
\begin{center}
\begin{tabular}{|c||c|c|c|c|}
\hline \hline
$i$ &   $P_i$ & $\tilde P_i$ & $  P_i^\sigma  $ &  $ \tilde P_i^\sigma  $  \\
\hline \hline
5 &   $(0,21)$ & $(0,28)$ & $ \tilde P_5 $ & $ P_5$ \\
6 &    $(3,9)$ & $(3,44)$ & $\tilde P_6$ & $ P_6$ \\
7 &   $(4,26)$ & $(4,33)$ & $\tilde P_7 $ & $P_7$ \\
\hline
8 &   $(7,5)$ & $(7,44)$ & $(42,5)$ & $(42,9)$ \\
9 &   $(8,22)$ & $(8,26)$ & $(43,29)$ & $(43,33)$ \\
10 &   $(11,5)$ & $(11,47)$ & $(46,5)$ & $(46,12)$ \\
11 &  $(12,6)$ & $(12,45)$ & $(47,6)$ & $(47,10)$  \\
12 &   $(13,14)$ & $(13,29)$ & $(48,35)$ & $(48,22)$ \\
13 &   $(14,8)$ & $(14,34)$ & $(35,43)$ & $(35,27)$ \\
14 &   $(15,13)$ & $(15,28)$ & $(36,48)$ & $(36,21)$ \\
15 &   $(16,17)$ & $(16,23)$ & $(37,38)$ & $(37,30)$ \\
16 &   $(17,0)$ & $(17,39)$ & $(38,0)$ & $(38,18)$ \\
17 &   $(18,4)$ & $(18,41)$ & $(39,4)$ & $(39,20)$ \\
18 &   $(19,9)$ & $(19,28)$ & $(40,44)$ & $(40,21)$ \\
19 &   $(20,12)$ & $(20,31)$ & $(41,47)$ & $(41,24)$ \\
20 &   $(22,4)$ & $(22,30)$ & $(29,4)$ & $(29,23)$ \\
21 &   $(25,6)$ & $(25,32)$ & $(32,6)$ & $(32,25)$ \\
22 &   $(27,7)$ & $(27,22)$ & $(34,42)$ & $(34,29)$ \\
\hline \hline
\end{tabular}
\bigskip
\caption{$\FF_{49}$-rational points of the curve $\calC$ 
(which are not $\FF_7$-rational); $P^\sigma$ denotes conjugate to $P$ with
respect to the action of the Frobenius automorphism.}
\label{tab:PunktyF49}
\end{center}
\end{table}

\begin{table}[!hp]
\begin{center}
\begin{tabular}{|r||r|c||r|c|} 
 \hline \hline 
$n$  & \hspace{1cm} $[n D_1]_r$ \hspace{1cm}   &
$g_0(n)$ & \hspace{.5cm} $[n D_1 + D_4 ]_r$ \hspace{1cm} & $g_1(n)$ \\
\hline \hline
0 &    $  0 \;$             & 1
	&$ (6,4) - \;\infty\:  $  & 1\\
1 &  $  \boldsymbol{(1,1)-  \;\infty\:} $     &  1
	&$ (1,1)+(6,4)- 2 \infty $ & $ \frac{5+5 u+3  {u^2}+v}{6+4  u+{u^2}} $	\\
2 & $ (1,1)+(1,1)- 2 \infty $  &  $\frac{u+5  {u^2}+v}{{{(2+u)}^2}} $  
	& $ (12,45)+(47,10)- 2 \infty $  &  $\frac{1+5  {u^2}+v}{2+5  u+{u^2}} $  \\
3 & $ (5,6)+(5,6)- 2 \infty $   &  $\frac{1+u+4  {u^2}+v}{(2+u) (5+u)} $ 
	& $ (15,28)+(36,21)- 2 \infty  $   & $\frac{6  {u^2}+v}{2+{u^2}} $ \\
4 & $ (2,3)+(5,3)- 2 \infty $   &  $\frac{2+4  {u^2}+v}{5+4  u+{u^2}} $ 
	& $ (7,44)+(42,9)- 2 \infty $   & $ \frac{5+u+v}{4+6  u+{u^2}} $  \\
5 & $ (19,9)+(40,44)- 2 \infty $   &   $\frac{4 u+2  {u^2}+v}{5+5  u+{u^2}} $
	& $ (11,5)+(46,5)- 2 \infty $   & $\frac{6+6 u+{u^2}+v}{3+6  u+{u^2}} $ \\
6 & $ (22,4)+(29,4)- 2 \infty $   &  $\frac{5+2  u+6  {u^2}+v}{(2+u)  (5+u)} $ 
	& $ (18,41)+(39,20)- 2 \infty $   & $\frac{5+3  u+5  {u^2}+v}{5+3  u+{u^2}} $  \\
7 & $ (2,3)+(5,6)- 2 \infty $      & $\frac{5+6 u+2  {u^2}+v}{5+2  u+{u^2}} $
	& $ (16,17)+(37,38)- 2 \infty $   & $\frac{5+4 u+4  {u^2}+v}{3+u+{u^2}} $  \\
8 & $ (27,22)+(34,29)- 2 \infty $      &$\frac{1+3  u+2  {u^2}+v}{1+{u^2}} $
	& $ (17,39)+(38,18)- 2 \infty $   & $\frac{3+2  u+{u^2}+v}{(5+u)  (6+u)} $   \\
9 & $ (14,34)+(35,27)- 2 \infty $      &$\frac{1+5 u+v}{(1+u)  (5+u)} $
	& $ (1,5)+(2,2)- 2 \infty $   &  $6+u $ \\
10 & $ (2,2)+(6,4)- 2 \infty $     &$\frac{3+5  u+5  {u^2}+v}{{{(5+u)}^2}} $
	& $ \boldsymbol{(2,2)-  \;\infty \:}$   & $1$ \\
11 & $ (2,3)+(2,3)- 2 \infty $     & $\frac{6+u+6 {u^2}+v}{3+2  u+{u^2}} $
	& $ (1,1)+(2,2)- 2 \infty $  &  $\frac{4+2  {u^2}+v}{3+5  u+{u^2}} $  \\
12 & $ (13,14)+(48,35)- 2 \infty $     & $\frac{3+6  u+4  {u^2}+v}{2+2  u+{u^2}} $
	& $ (8,22)+(43,29)- 2 \infty $    & $\frac{2+4  u+v}{(2+u) (6+u)} $\\
13 & $ (20,12)+(41,47)- 2 \infty $     & $\frac{5 u+{u^2}+v}{(1+u)  (2+u)} $
	& $ (1,5)+(5,3)- 2 \infty $   & $6+u $   \\
14 & $ (5,3)+(6,4)- 2 \infty $     &$ \frac{6+5  u+2  {u^2}+v}{6+6  u+{u^2}} $ 
	& $ (5,3) - \;\infty \:$   & $1$   \\
15 & $ (25,32)+(32,25)- 2 \infty $     &$\frac{5+{u^2}+v}{6+6 u+{u^2}} $
	& $ (1,1)+(5,3)- 2 \infty $  &  $\frac{u+5  {u^2}+v}{(2+u) (6+u)} $  \\
16 & $ (25,6)+(32,6)- 2 \infty $     &$\frac{6+5  u+2  {u^2}+v}{(1+u)  (2+u)} $
	& $ (1,5)+(5,6)- 2 \infty $  & $6+u $    \\
17 & $ (5,6)+(6,4)- 2 \infty $     &$\frac{5 u+{u^2}+v}{2+2  u+{u^2}} $
	& $ (5,6) -  \;\infty \:$   &  $1$  \\ 
18 & $ (20,31)+(41,24)- 2 \infty $     &$\frac{3+6  u+4  {u^2}+v}{3+2  u+{u^2}} $ 
	& $ (1,1)+(5,6)- 2 \infty $   &  $\frac{2+4  u+v}{3+5 u+{u^2}} $ \\
19 & $ (13,29)+(48,22)- 2 \infty $     &$\frac{6+u+6 {u^2}+v}{{{(5+u)}^2}} $
	& $ (8,26)+(43,33)- 2 \infty $   &  $\frac{4+2  {u^2}+v}{(5+u)  (6+u)} $\\
20 & $ (2,2)+(2,2)- 2 \infty $     &$\frac{3+5  u+5  {u^2}+v}{(1+u)  (5+u)} $ 
	& $ (1,5)+(2,3)- 2 \infty $  &   $6+u $ \\
21 & $ (2,3)+(6,4)- 2 \infty $     &$\frac{1+5 u+v}{1+{u^2}} $
	& $ (2,3)- \;\infty\: $    & $ 1 $\\
22 & $ (14,8)+(35,43)- 2 \infty $     &$\frac{1+3  u+2  {u^2}+v}{5+2  u+{u^2}} $
	& $ (1,1)+(2,3)- 2 \infty $   & $\frac{3+2  u+{u^2}+v}{3+u+{u^2}}$  \\
23 & $ (27,7)+(34,42)- 2 \infty $     &$\frac{5+6 u+2  {u^2}+v}{(2+u)  (5+u)} $
	& $ (17,0)+(38,0)- 2 \infty $   &  $\frac{5+4 u+4  {u^2}+v}{5+3  u+{u^2}} $\\
24 & $ (2,2)+(5,3)- 2 \infty $     &$\frac{5+2  u+6  {u^2}+v}{5+5  u+{u^2}} $
	& $ (16,23)+(37,30)- 2 \infty $  &  $\frac{5+3  u+5  {u^2}+v}{3+6  u+{u^2}}$  \\
25 & $ (22,30)+(29,23)- 2 \infty $     &$\frac{4 u+2  {u^2}+v}{5+4  u+{u^2}} $ 
	& $ (18,4)+(39,4)- 2 \infty $  &   $\frac{6+6  u+{u^2}+v}{4+6  u+{u^2}}$\\
26 & $ (19,28)+(40,21)- 2 \infty $     &$\frac{2+4  {u^2}+v}{(2+u)  (5+u)} $
	& $ (11,47)+(46,12)- 2 \infty $  &  $\frac{5+u+v}{2+{u^2}}$  \\
27 & $ (2,2)+(5,6)- 2 \infty $     &$\frac{1+u+4 {u^2}+v}{{{(2+u)}^2}}$ 
	& $ (7,5)+(42,5)- 2 \infty $   & $\frac{6  {u^2}+v}{2+5 u+{u^2}} $ \\
28 & $ (5,3)+(5,3)- 2 \infty $     &$\frac{u+5u^2+v}{(6+u)^2}$
	& $ (15,13)+(36,48)- 2 \infty $   & $\frac{1+5  {u^2}+v}{6+4  u+{u^2}}$ \\
29 & $ (1,5)+(1,5)- 2 \infty $     & $(6+u)$  
	& $ \boldsymbol{(12,6)+(47,6)- 2 \infty} $  &   $\frac{5+5  u+3 {u^2}+v}{(1+u)  (6+u)}  $   \\
30 & $ \boldsymbol{(1,5)-  \;\infty \:}$     &$(6+u)$
	& $ (1,5)+(6,4)- 2 \infty $   & $6+u $  \\
\hline \hline

\end{tabular}
\bigskip
\caption{The group $J(\calC;\FF_7)$ as the simple sum of its cyclic subgroups;
$[X]_r$ denotes the reduced representsnt of the equivalence class of the divisor $X$;
\quad $g_0(n), g_1(n)$ --- transition functions (see main text).
}
\label{tab:jakobianN}
\end{center}
\end{table}

The full description of the group $J(\calC;\FF_7)$
is given in Table~\ref{tab:jakobianN}, where we have choosen as a reference point 
$A_0$ the infinity point $\infty$. 
The divisor $D_1=P_1 -\infty$ generates the subgroup of order $31$ and the
divisor $D_4=P_4 -\infty$ generates the subgroup of order $2$.
For $n\in\{ 0,1,\dots,30 \}$ and $ m\in\{ 0,1 \}$
we present the reduced representants of elements $[nD_1 + mD_4]_r$ of 
$J(\calC;\FF_7)$ 
 and the transition functions $g_m(n)$ which are given by the equation
$$[nD_1 + mD_4]_r + D_1= (g_m(n))+ [(n+1)D_1 + mD_4]_r,$$
and normalized (numerators and denominators are monic polynomials).
We will use them in the construction below.

\subsection{Construction of the wave and $\tau$ functions}

In order to find a solution of the discrete KdV equation let us fix the
following points of the curve $\calC$:  
\begin{equation*} 
A_0=\infty, \quad A_1=(2,2), \quad A_2=(1,5), \quad A_3=(1,1),
\end{equation*}  
with the uniformizing parameters
$ t_0=u^2/v, \quad t_1=u-2, \quad t_2=t_3=u-1$,  
and
\begin{equation*}
B_1=(12,6), \quad B_2=(47,6).
\end{equation*}
Then
\begin{gather*}
A_1 - A_0 \sim 10D_1 + D_4, \qquad  
A_2-A_0 \sim -D_1 \sim 30 D_1, \qquad A_3 - A_0 \sim D_1, \\
B_1+B_2 - 2 A_0 \sim 29D_1 + D_4,
\end{gather*}
and the points $X_1(n_1,n_2,n_3)$ and $X_2(n_1,n_2,n_3)$, where the wave 
function $\psi(n_1,n_2,n_3)$ 
has additional zeros (here $X_i$ {\bf can} be $\infty$) 
can be found from Table~\ref{tab:jakobianN} and 
\begin{equation} \label{eqn:X_iTOn}
X_1(n_1,n_2,n_3) + X_2(n_1,n_2,n_3) - 2 \infty = [n D_1 + m D_4]_r ,
\end{equation}
where $n\in\{ 0,1,\dots,30 \}$ and $ m\in\{ 0,1 \}$ are given by 
\begin{eqnarray}\label{eqn:n_iTOn}
&n \equiv  29-(n_3-n_2+10 n_1)  \mod 31, \\ 
&m \equiv  1-n_1  \mod 2. \label{eqn:n_iTOm}
\end{eqnarray}
\begin{Rem}
The choice of the infinity point $\infty$ as $A_0$ is a violation of 
the assumption of general position of points used in the construction 
($\infty$ is the Weierstrass point of the curve $\calC$).  
This will not destroy the construction but in some situations, which we will
point out, will affect uniqueness of the wave function. We remark that such a
choice is indispensable in reduction of the method from the discrete KP 
equation to the discrete KdV equation (see, for example \cite{KWZ,BD-KP}).
\end{Rem}
The key idea in constructing of the wave function is to express $\psi(n_1,n_2,n_3)$ 
for any parameters from $(n_1,n_2,n_3)\in\ZZ^3$ by a set of functions related to $J(\calC;\KK)$
(transition functions and few auxiliary functions).
Let us introduce functions $h_1$ and $h_4$ corresponding to generators
of the two cyclic subgroups of $J(\calC;\FF_7)$. The function $h_1$ with the
divisor $31 D_1\sim 0$ and normalized at the infinity point is equal to
\begin{equation*}
h_1 = \prod_{i=0}^{30} g_0(i), 
\end{equation*}
and reads
\begin{multline*}
h_{1} =
1+2 u+{u^2}+4 {u^3}+3 {u^5}+{u^6}+ 3 {u^7}+ {u^8}+ 4 {u^9}+\\ 4 {u^{10}}+
2 {u^{11}}+5 {u^{12}}+2 {u^{13}}+ 4 {u^{14}}+
3 {u^{15}}+\big(5 u+2 {u^2}+ \\ 5 {u^3}+ 4 {u^5}+6 {u^6}+4 {u^7}+3
{u^9}+5 {u^{10}}+5 {u^{11}}+4 {u^{12}}+{u^{13}}\big) v ,
\end{multline*}
where we also used equation of the curve \eqref{eq:curve} to reduce higher order
terms in $v$.
The normalized function $h_4$ with the divisor $2D_4\sim 0$ is 
\[ h_4 = u-6.
\]
Let us introduce other auxilliary function $f_1$ and $f_2$
to factorise the zeros at $A_1$ and $A_2$ of the wave
function.
Notice that 
\[ (2,2) + 21 (1,1) + (6,4) - 23 \infty \sim 0,
\]
which implies that there exists a polynomial function on 
$\calC$ with simple zero at $A_1$ and other zeros in the distinguished (by our
choice of description of $J(\calC;\FF_7)$) points $(1,1)$ and $(6,4)$.
Define $f_1$ as the unique such function normalized at the infinity point
$\infty$, then
\begin{multline*}
f_1 =1+5 u+{u^2}+4 {u^4}+6 {u^5}+4 {u^6}+4 {u^7}+3 {u^8}+4 {u^9}+ \\6 {u^{11}}+
\big(6+4 u+2 {u^2}+5 {u^3}+6 {u^4}+6 {u^6}+{u^7}+{u^8}+{u^9}\big) v.
\end{multline*}
The zeros at $A_2$ can be factorised using function $$f_2=(u-1).$$

Uniqueness of the wave function $\psi$ implies that it
can be decomposed as follows
\begin{equation} \label{eqn:psi-W}
\psi(n_1,n_2,n_3)=\frac{f_1^{n_1} f_2^{n_2} }{ h_{1}^p h_4^q} W(m_1,m_2),
\end{equation}
where new variables  $m_1$ i $m_2$ are given by 
\begin{eqnarray} 
\label{eqn:m-n1}
 21 n_1 - n_2 + n_3 &=& 31p-m_1, \quad m_1\in\{ 0,1,\dots,30 \}, \\
\label{eqn:m-n2}
 n_1 &=& 2q-m_2, \quad m_2\in\{ 0,1 \}, 
\end{eqnarray} 
and the function $W(m_1,m_2)$ has the divisor
\begin{equation} \label{eqn:Y_iTOm_i}
m_1 D_1 + m_2 D_4 +Y_1(m_1,m_2) + Y_2(m_1,m_2) 
-(12,6) - (47,6) .
\end{equation}
The additional zeros 
$$Y_1(m_1,m_2)+Y_2(m_1,m_2)=X_1(n_1,n_2,n_3)+X_2(n_1,n_2,n_3),$$ 
can be found by projection from
$n$-variables into the $m$-variables.

To find  the functions $W(m_1,m_2)$ for all $m_1\in\{ 0,1,\dots,30 \}$ 
and $ m_2\in\{ 0,1 \}$ let us notice that $W(0,0)=1$ and $W(0,1)$ 
is given by 
$$ (W(0,1)) = D_4 +(1,5)+ (1,5) - (12,6) - (47,6), $$
and hence can be written in a form 
$$ W(0,1)= \frac{2+3u+4u^2+v}{6+4u+u^2}.$$
 Define the multipliers $w_{m_2}(m_1)$ as follows
\begin{equation*}
    W(m_1,m_2) = w_{m_2}(m_1) W(m_1-1,m_2),
\end{equation*}
for $m_1\in\{ 0,1,\dots,30 \}$ and $m_2\in\{0,1\}$.
Equations \eqref{eqn:X_iTOn}-\eqref{eqn:n_iTOm} and \eqref{eqn:m-n1}-\eqref{eqn:Y_iTOm_i} gives 
$$ w_{m_2}(m_1)=g_m(n),$$
where
$$ m_2=1-m \mod 2, \qquad  m_1=29-n \mod 31.$$ 
Setting $w_{m_2}(0)=W(0,m_2)$ we obtain 
\begin{equation*}
	W(m_1,m_2)= \prod_{i=0}^{m_1} w_{m_2}(i). 
\end{equation*}
Together with factorisation \eqref{eqn:psi-W} it gives the wave function
$\psi$ for all $(n_1,n_2,n_3)\in \ZZ^3$. 

\begin{Rem}
For $(m_1,m_2)=(29,1)$ we have $X_1=X_2=\infty$. Because the infinity point
$\infty$ is the Weierstrass point of order two, there exist functions with
divisor of poles equal to $2\infty$. This means that $\psi$ is not uniqely 
determined in this case. However it is natural to keep the divisor of $\psi$,
and therefore $\psi$ itself, exactly like it is given from the flow on $J(\calC;\KK)$.
Notice that because for $X_1=X_2=\infty$ we stay in non-generic case,
then this ambiguity does not affect construction of the $\tau$-function.   
\end{Rem}

\begin{figure}[!t]
\begin{center}
\leavevmode\epsfysize=5.6cm\epsffile{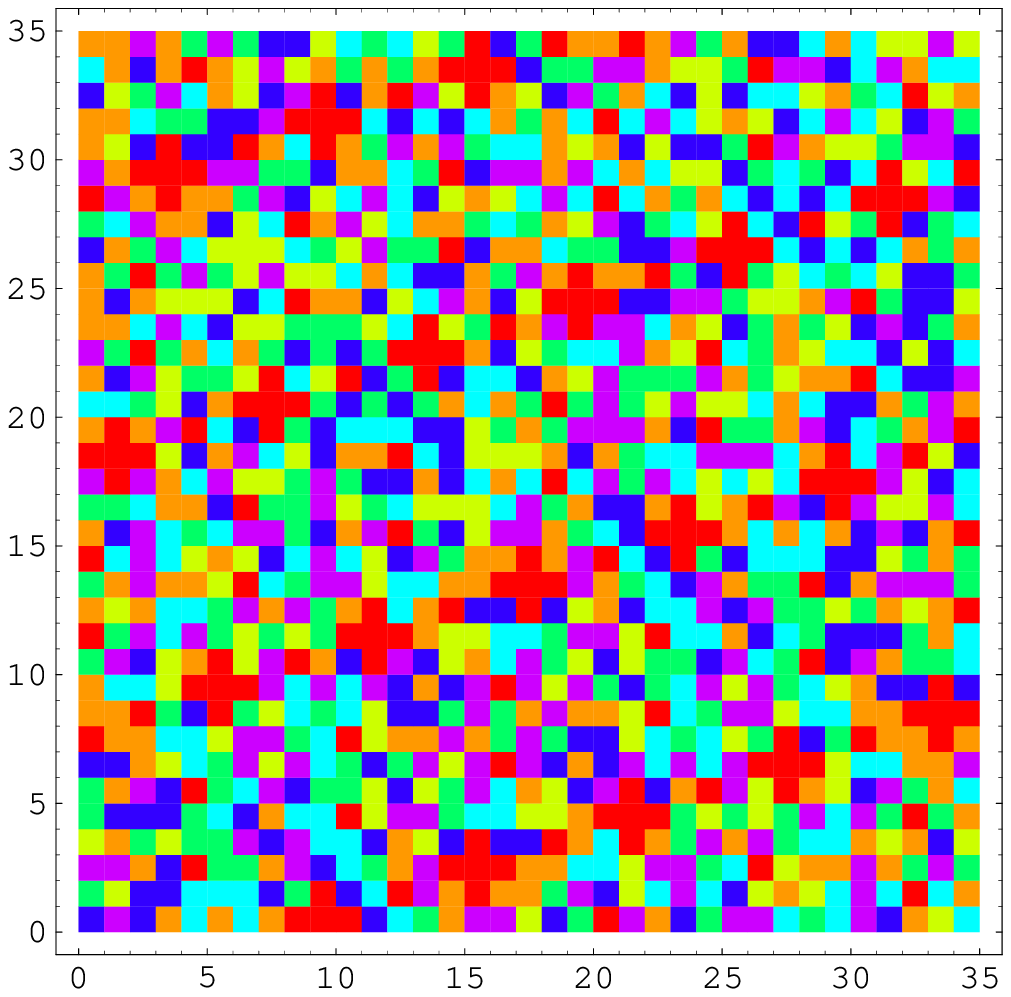} \hspace{0.5cm}
\leavevmode\epsfysize=5.6cm\epsffile{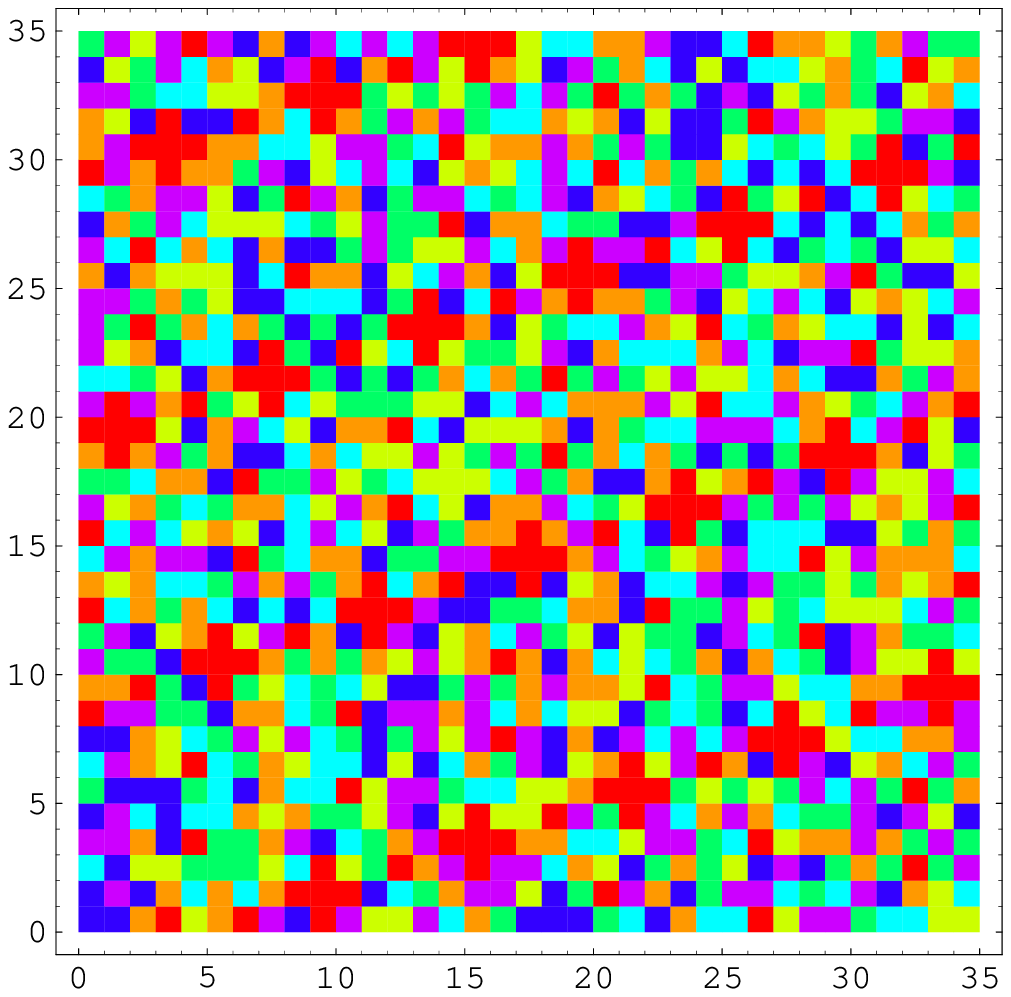}
\leavevmode\epsfysize=5.6cm\epsffile{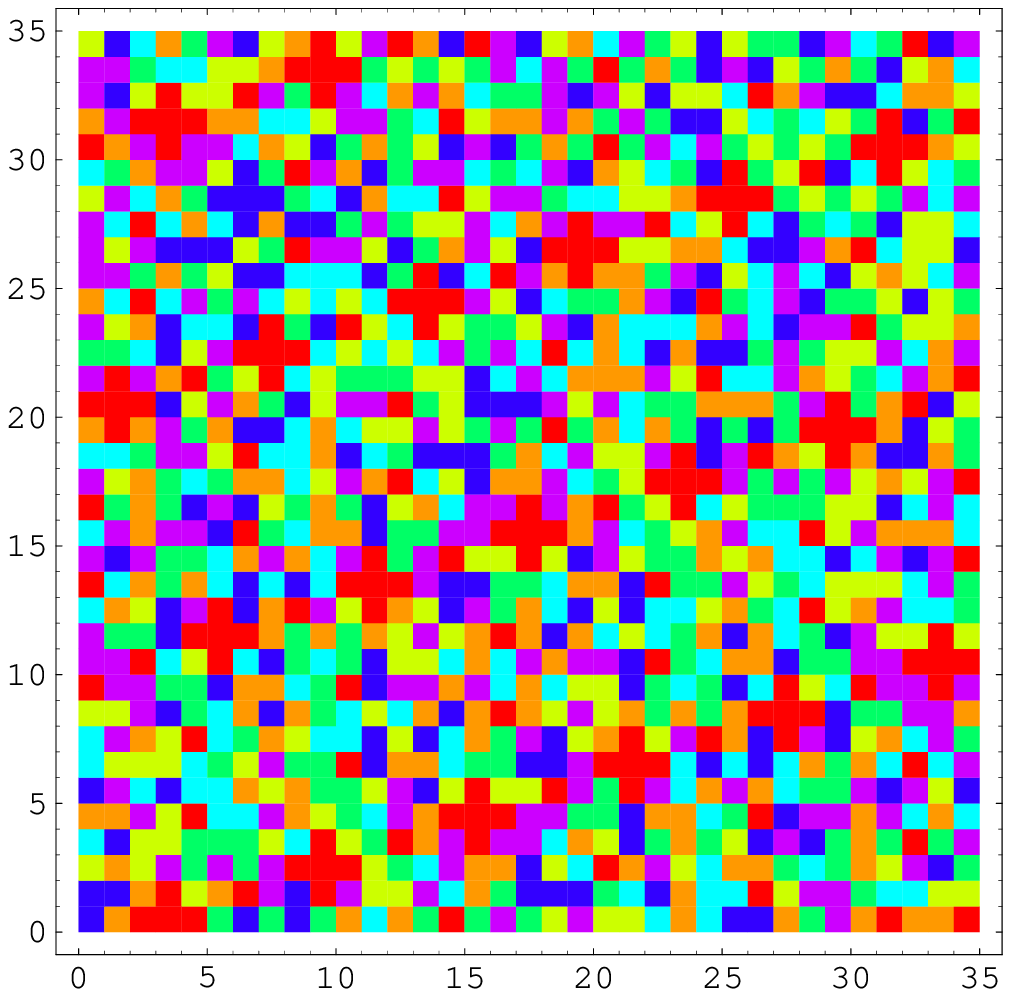} \hspace{0.5cm}
\leavevmode\epsfysize=5.6cm\epsffile{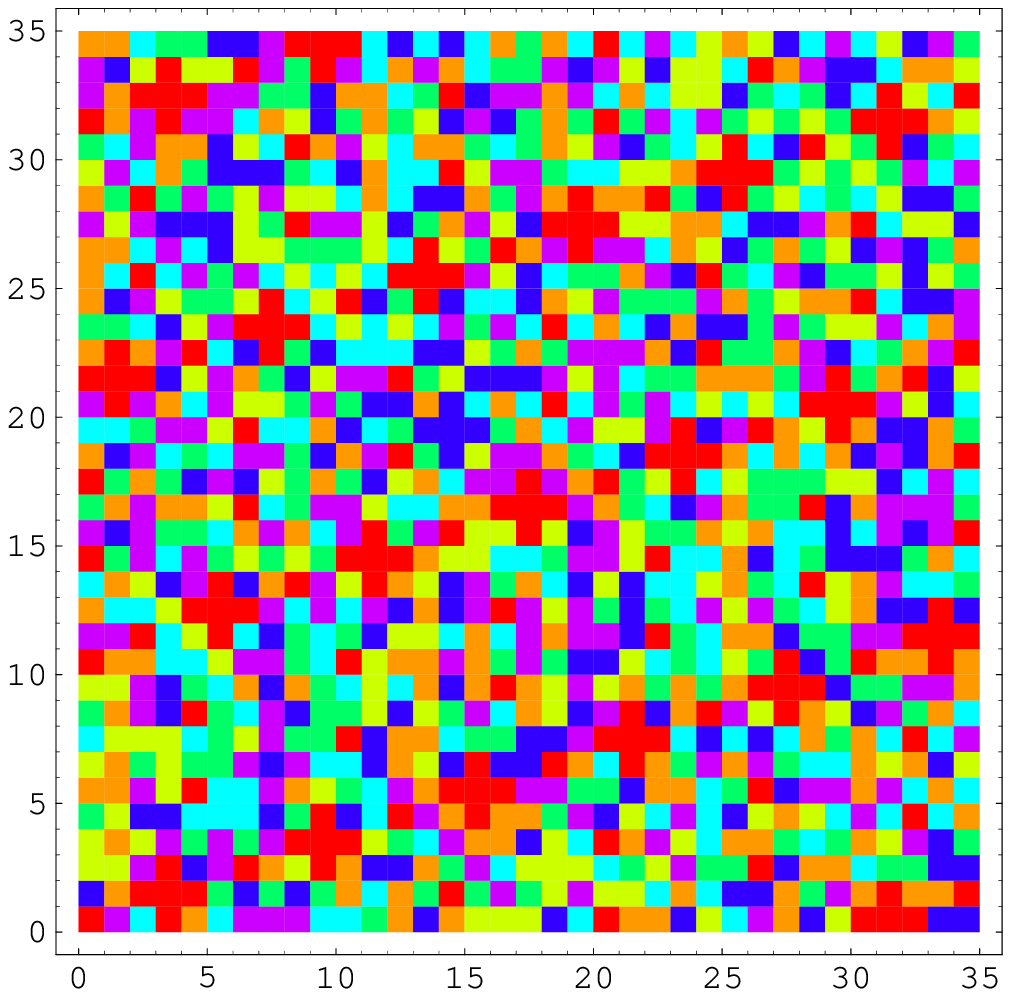} 

\leavevmode\epsfysize=0.35cm\epsffile{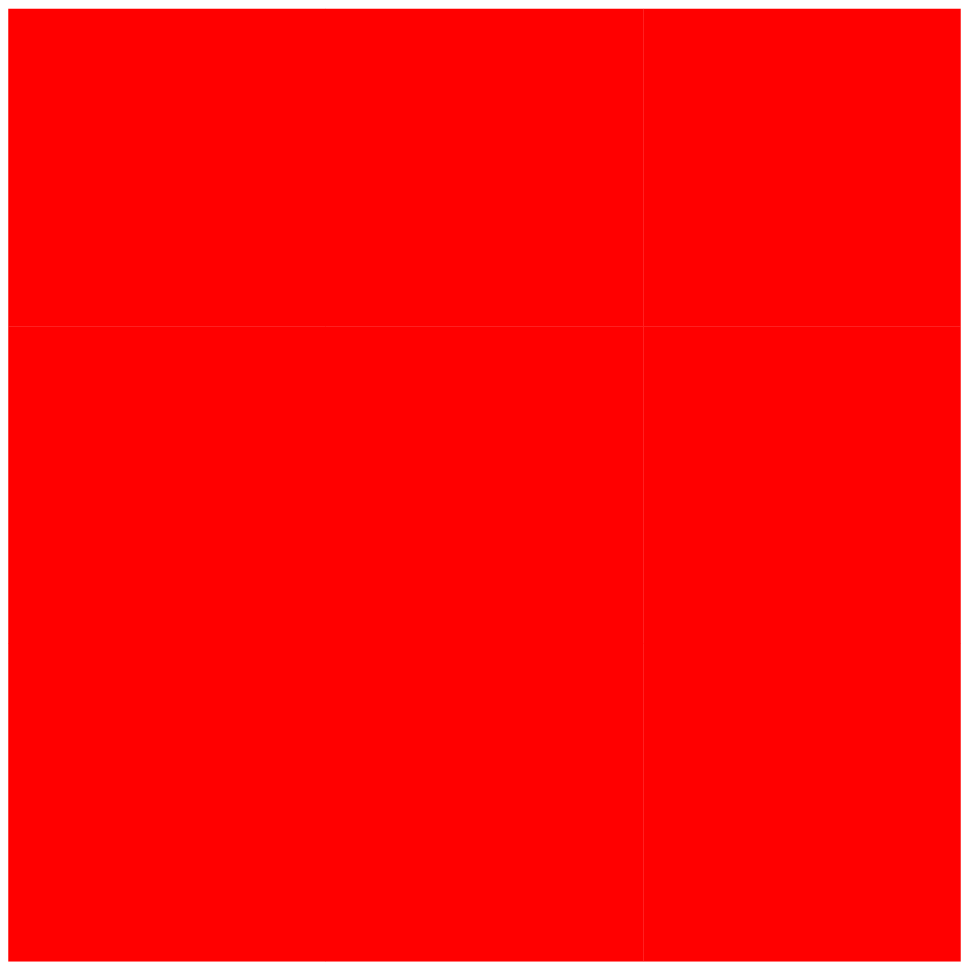} -- $(00)$, 
\leavevmode\epsfysize=0.35cm\epsffile{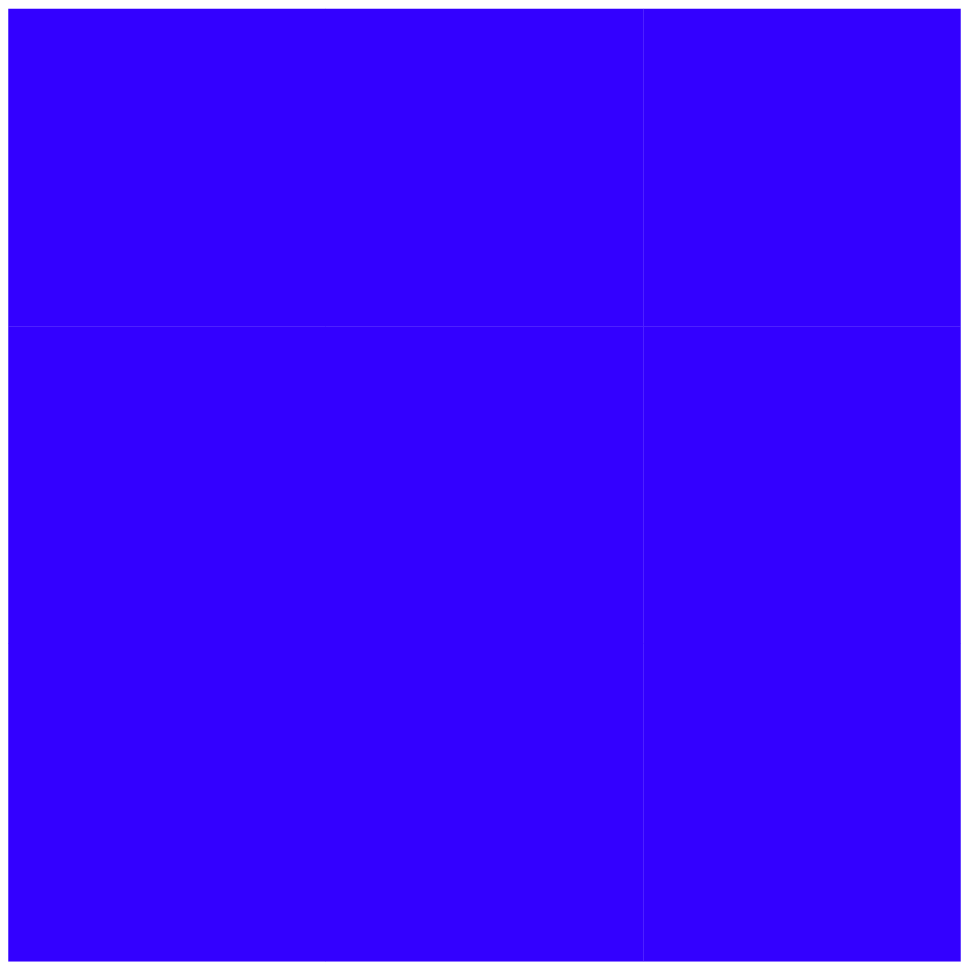} -- $(01)$, 
\leavevmode\epsfysize=0.35cm\epsffile{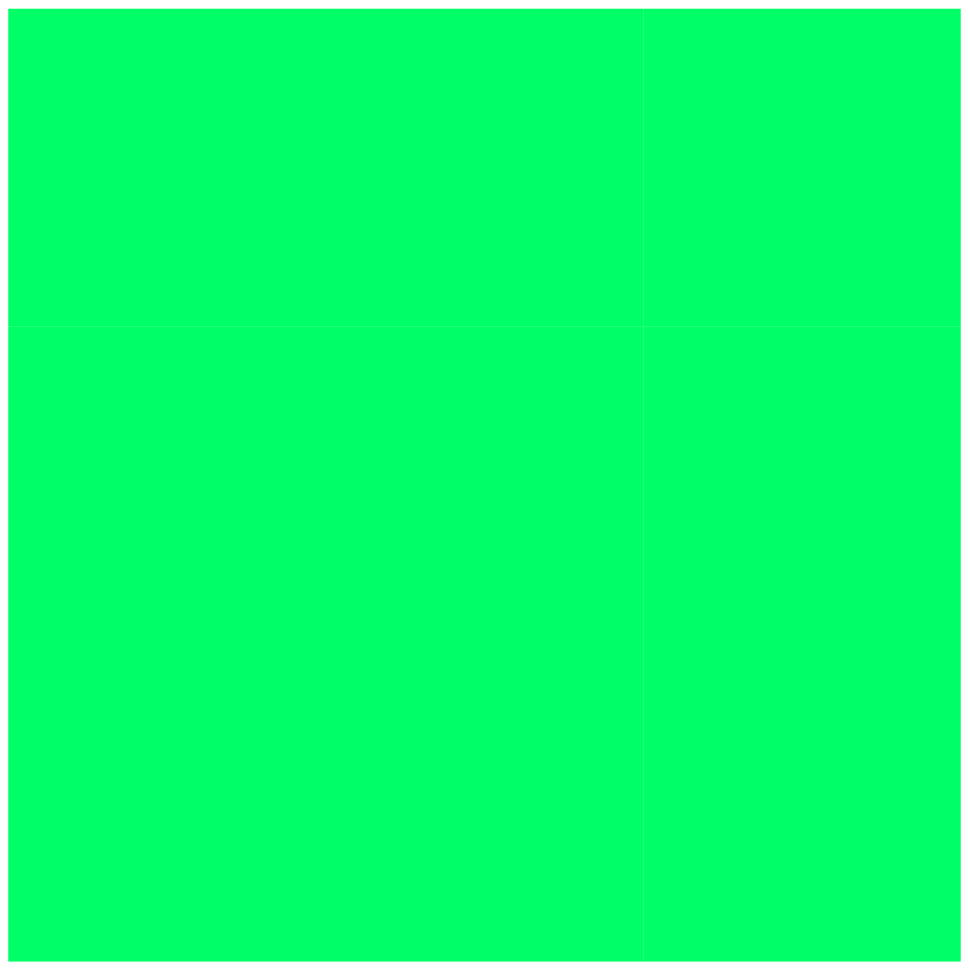} -- $(02)$,
\leavevmode\epsfysize=0.35cm\epsffile{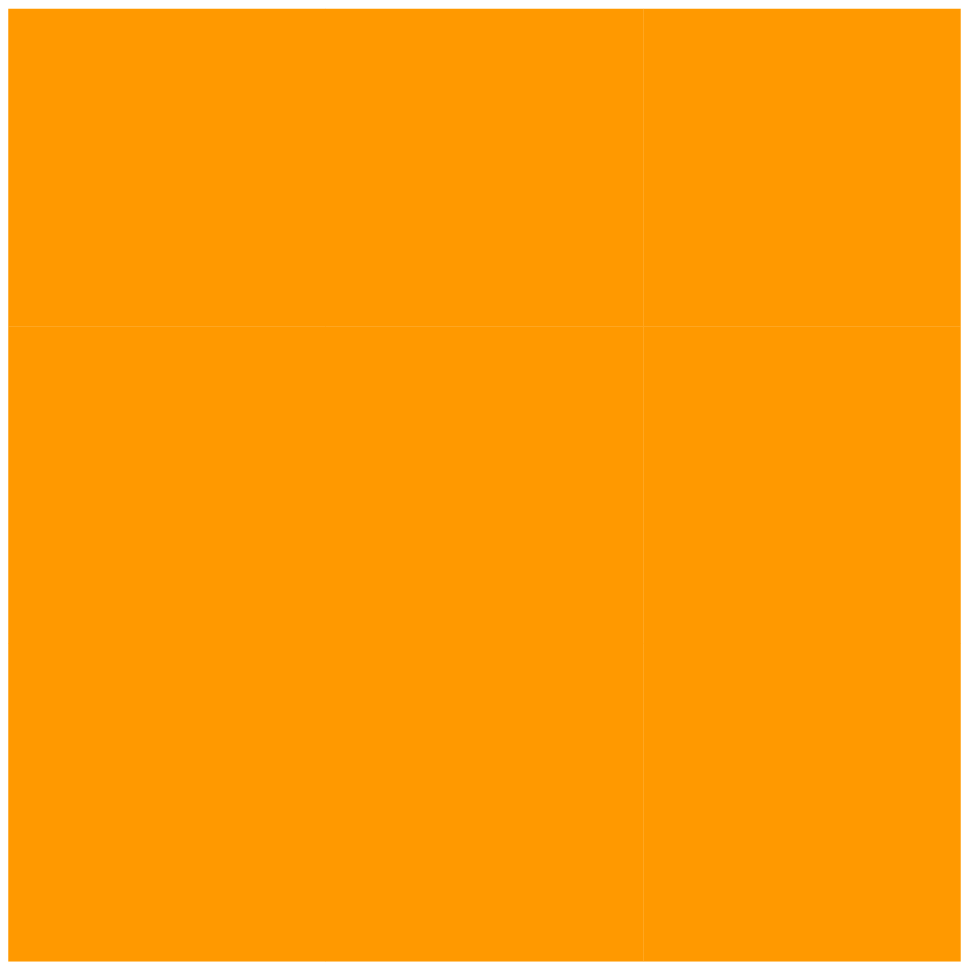} -- $(03)$, 
\leavevmode\epsfysize=0.35cm\epsffile{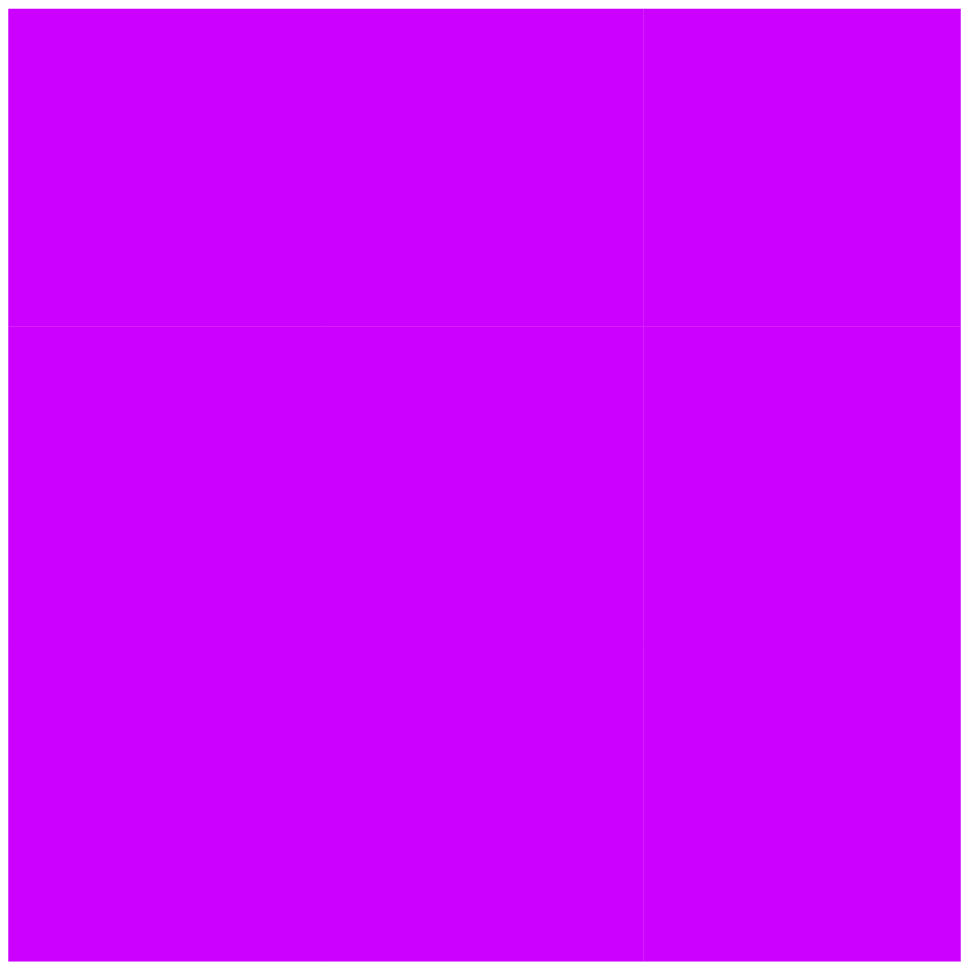} -- $(04)$, 
\leavevmode\epsfysize=0.35cm\epsffile{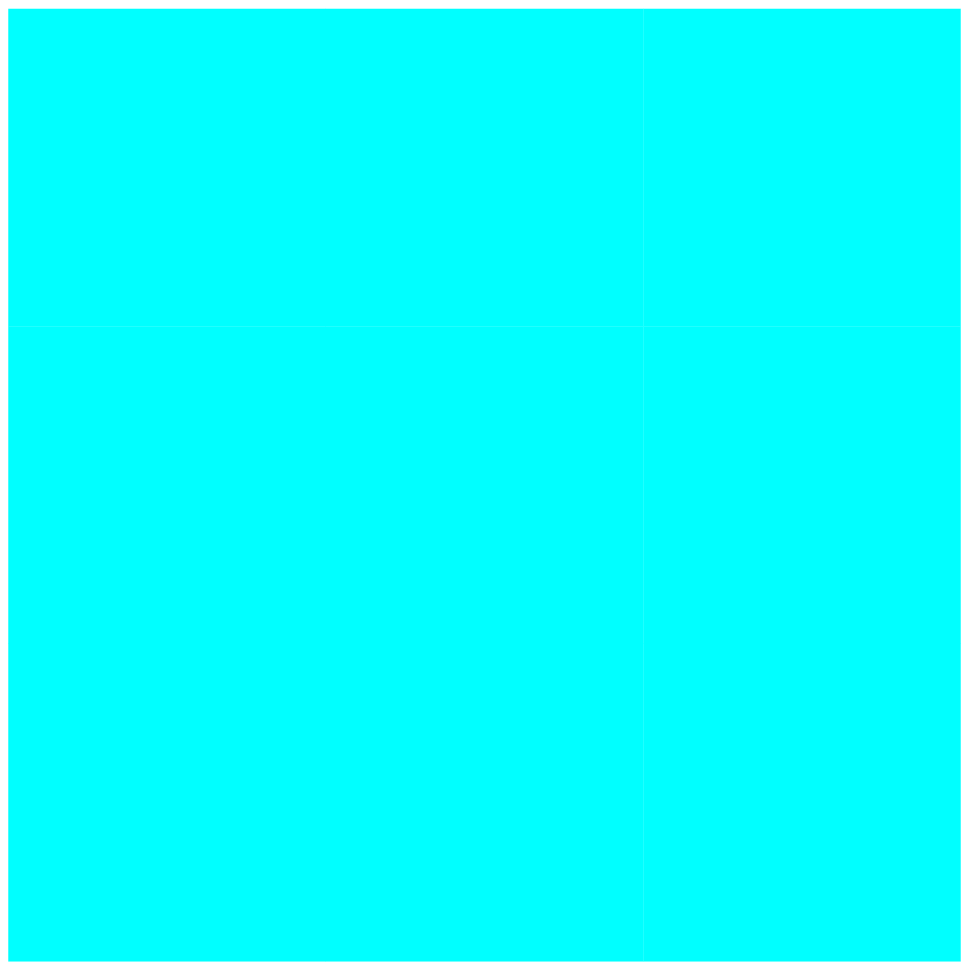} -- $(05)$, 
\leavevmode\epsfysize=0.35cm\epsffile{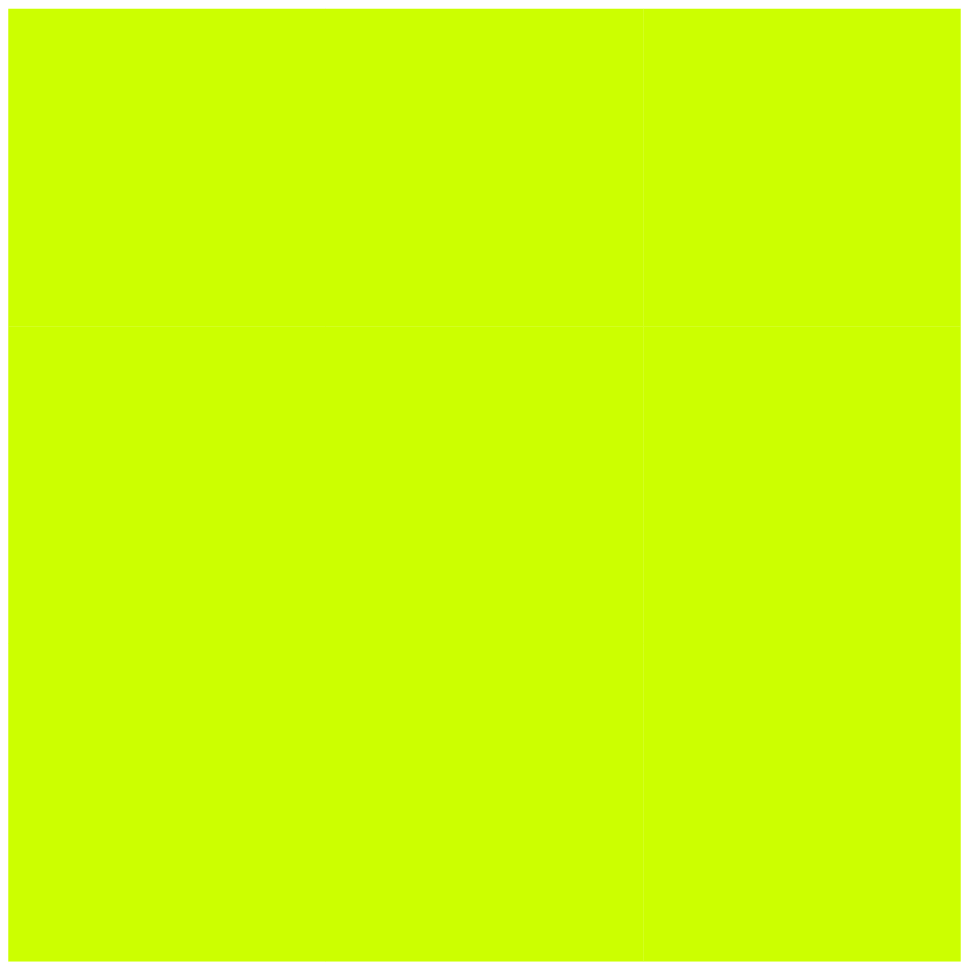} -- $(06)$.

\end{center}
\caption{
The $\FF_7$-valued solution of the discrete KP equation out of genus $g=2$ hyperelliptic curve $\calC$; 
$n_2=-1,0,1,2$ for subseqent figures, $n_1=0,1,\ldots,34$ (horizontal axis),
 $n_3=0,1,\ldots,34$ (vertical axis).
}
\label{fig:g2aKPforKdV}
\end{figure}

The coefficients ${\zeta_0}^{(k)}(n_1,n_2,n_3)$, $k=1,2,3$, of expansion of the
wave function can be obtained from factorisation \eqref{eqn:psi-W}
and are given by
\begin{eqnarray}
\label{eqn:z-z1}
  (2,2):\quad \zeta_0^{(1)} (n_1,n_2,n_3)&=& 
 6^{n_1} 5^{q} 4^p  W(m_1,m_2)|_{t_1=0} ,\\
\label{eqn:z-z2}
 (1,5):\quad \zeta_0^{(2)} (n_1,n_2,n_3)&=& 
 2^{n_1} 4^q W(m_1,m_2)|_{t_2=0} ,\\
\label{eqn:z-z3}
 (1,1):\quad \zeta_0^{(3)} (n_1,n_2,n_3)&=& 
6^{p} 4^q \frac{W(m_1,m_2)}{t_3^{m_1}}|_{t_3=0}. 
\end{eqnarray}
Using definition of the $\tau$ function for nonzero $\rho_i$, i.e. 
equation~\eqref{eq:tau-def},
and putting $\tau=0$ for points related with nongeneric case
we obtain a solution of the discrete KP
equation~\eqref{eq:tauKP} taking value in the finite field $\FF_7$. 
This $\tau$-function is presented in Figure~\ref{fig:g2aKPforKdV}. 

\begin{figure}[!t]
\begin{center}

\leavevmode\epsfysize=10.5cm\epsffile{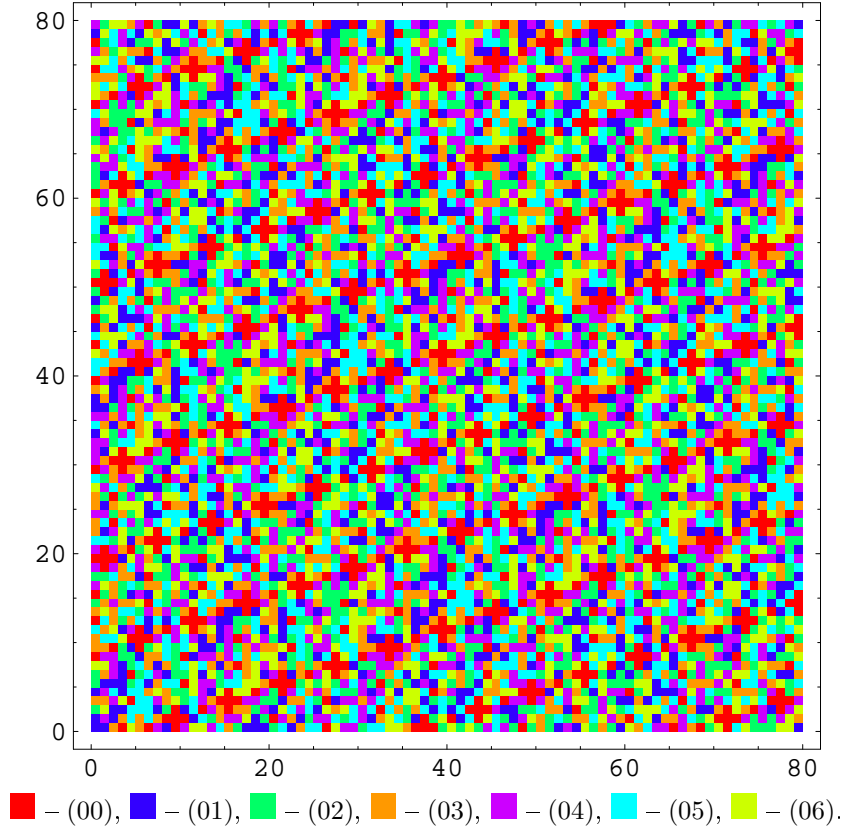} 

\leavevmode\epsfysize=0.35cm\epsffile{f7-0.eps} -- $(00)$, 
\leavevmode\epsfysize=0.35cm\epsffile{f7-1.eps} -- $(01)$, 
\leavevmode\epsfysize=0.35cm\epsffile{f7-2.eps} -- $(02)$,
\leavevmode\epsfysize=0.35cm\epsffile{f7-3.eps} -- $(03)$, 
\leavevmode\epsfysize=0.35cm\epsffile{f7-4.eps} -- $(04)$, 
\leavevmode\epsfysize=0.35cm\epsffile{f7-5.eps} -- $(05)$, 
\leavevmode\epsfysize=0.35cm\epsffile{f7-6.eps} -- $(06)$. 

\end{center}
\caption{
The $\FF_7$-valued solution of the discrete KdV equation out of genus $g=2$ 
hyperelliptic curve $\calC$; 
$n_1=0,1,\ldots,79$ (horizontal axis),
 $n_3=0,1,\ldots,79$ (vertical axis), $(n_2=0)$.
}
\label{fig:g2KdVbig}
\end{figure}

To obtain a $\tilde \tau$-function being a solution of the 
discrete KdV equation we use the formula~\eqref{eq:tau-tau-tylda}.
For our settings we have $\delta_2=\delta_3=1$, so finally 
$$ \tilde \tau= \tau (-1)^{-n_3(n_3-1)/2}.$$  
The $\tilde \tau$-function for dKdV is presented 
in Figure~\ref{fig:g2KdVbig}.

\section*{Acknowledgments}
Author would like to thank prof. Adam Doliwa for "multidimensional" help.

The paper was partially supported by the KBN grant 2~P03B~12622.

\bibliographystyle{amsplain}

\begin{thebibliography}{10}

\bibitem{BBEIM}
E.~D. Belokolos, A.~I. Bobenko, V.~Z. Enol'skii, A.~R. Its, and V.~B. Matveev,
  \emph{Algebro-geometric approach to nonlinear integrable equations},
  Springer-Verlag, Berlin, 1994.

\bibitem{Bial-phd}
M.~Bia{\l}ecki, \emph{Methods of algebraic geometry over finite fields in
  construction of integrable cellular automata}, {PhD} dissertation, Warsaw
  University, Institute of Theoretical Physics, 2003, (in Polish).

\bibitem{BD-hyp}
M.~Bia{\l}ecki and A.~Doliwa, \emph{Algebro-geometric solution of the d{KP}
  equation over a finite field out of a hyperelliptic curve}, {\tt
  nlin.SI/0309071}.

\bibitem{BD-KP}
\bysame, \emph{The discrete {KP} and {KdV} equations over finite fields},
  Theor. Math. Phys. \textbf{137(1)} (2003), 1412--1418.

\bibitem{ArithGeom}
G.~Cornell and J.~H.~Silverman (eds.), \emph{Arithmetic geometry},
  Springer-Verlag, New York, 1986.

\bibitem{DBK}
A.~Doliwa, M.~Bia{\l}ecki, and P.~Klimczewski, \emph{The {H}irota equation over
  finite fields: algebro-geometric approach and multisoliton solutions}, J.
  Phys. A: Math. Gen. \textbf{36} (2003), 4827--4839.

\bibitem{HirotaKdV}
R.~Hirota, \emph{Nonlinear partial difference equations. {I}. {A} difference
  analogue of the {K}orteweg-de {Vries} equation}, J. Phys. Soc. Jpn.
  \textbf{43} (1977), 1424--1433.

\bibitem{Hirota}
\bysame, \emph{Discrete analogue of a generalized {Toda} equation}, J. Phys.
  Soc. Jpn. \textbf{50} (1981), 3785--3791.

\bibitem{Koblitz-Ang}
N.~Koblitz, \emph{Algebraic aspects of cryptography}, Springer, Berlin, 1998.

\bibitem{Krich-discr}
I.~M. Krichever, \emph{Algebraic curves and nonlinear difference equations},
  Uspiekhi Mat. Nauk \textbf{33:4} (1978), 215--216.

\bibitem{KWZ}
I.~M. Krichever, P.~Wiegmann, and A.~Zabrodin, \emph{Elliptic solutions to
  difference non-linear equations and related many body problems}, Commun.
  Math. Phys. \textbf{193} (1998), 373--396.

\bibitem{Lang}
S.~Lang, \emph{Abelian varieties}, Interscience Publishers, Inc., New York,
  1958.

\bibitem{hiper-dodatekAng}
A.J. Menezes, Y.H. Wu, and R.J. Zuccherato, \emph{An elementary introduction to
  hipereliptic curves}, Appendix in \cite{Koblitz-Ang}.

\bibitem{Milne}
J.S. Milne, \emph{Jacobian varieties}, {C}hapter {VII} in \cite{ArithGeom}.

\bibitem{Moreno}
C.~Moreno, \emph{Algebraic curves over finite fields}, University Press,
  Cambridge, 1991.

\bibitem{Shafarevich}
I.~Shafarevich, \emph{Basic {A}lgebraic {G}eometry}, Springer-Verlag,
  Heidelberg, 1974.

\bibitem{Sticht}
H.~Stichtenoth, \emph{Algebraic function fields and codes}, Springer-Verlag,
  Berlin, 1993.

\end{thebibliography}

\providecommand{\bysame}{\leavevmode\hbox to3em{\hrulefill}\thinspace}
\providecommand{\MR}{\relax\ifhmode\unskip\space\fi MR }
\providecommand{\MRhref}[2]{%
  \href{http://www.ams.org/mathscinet-getitem?mr=#1}{#2}
}
\providecommand{\href}[2]{#2}

\end{document}